\documentstyle[graphicx]{./mn}

\newif\ifAMStwofonts

\def\mnras{MNRAS}

\def\aap{A\&A}

\ifoldfss
  \ifCUPmtlplainloaded \else
    \NewTextAlphabet{textbfit} {cmbxti10} {}
    \NewTextAlphabet{textbfss} {cmssbx10} {}
    \NewMathAlphabet{mathbfit} {cmbxti10} {} 
    \NewMathAlphabet{mathbfss} {cmssbx10} {} 
  \fi
  \ifAMStwofonts
    \ifCUPmtlplainloaded \else
      \NewSymbolFont{upmath} {eurm10}
      \NewSymbolFont{AMSa} {msam10}
      \NewMathSymbol{\upi}     {0}{upmath}{19}
      \NewMathSymbol{\umu}     {0}{upmath}{16}
      \NewMathSymbol{\upartial}{0}{upmath}{40}
      \NewMathSymbol{\leqslant}{3}{AMSa}{36}
      \NewMathSymbol{\geqslant}{3}{AMSa}{3E}

      \let\leq=\leqslant 
       
    \fi
  \fi
\fi 

\ifnfssone
  \newmathalphabet{\mathit}
  \addtoversion{normal}{\mathit}{cmr}{m}{it}
  \addtoversion{bold}{\mathit}{cmr}{bx}{it}
  \newmathalphabet{\mathbfit} 
  \addtoversion{normal}{\mathbfit}{cmr}{bx}{it}
  \addtoversion{bold}{\mathbfit}{cmr}{bx}{it}
  \newmathalphabet{\mathbfss} 
  \addtoversion{normal}{\mathbfss}{cmss}{bx}{n}
  \addtoversion{bold}{\mathbfss}{cmss}{bx}{n}
  \ifAMStwofonts
    \ifCUPmtlplainloaded \else
      \UseAMStwoboldmath
      \makeatletter
      \new@mathgroup\upmath@group
      \define@mathgroup\mv@normal\upmath@group{eur}{m}{n}
      \define@mathgroup\mv@bold\upmath@group{eur}{b}{n}
      \edef\UPM{\hexnumber\upmath@group}
      \new@mathgroup\amsa@group
      \define@mathgroup\mv@normal\amsa@group{msa}{m}{n}
      \define@mathgroup\mv@bold\amsa@group{msa}{m}{n}
      \edef\AMSa{\hexnumber\amsa@group}
      \makeatother
      \mathchardef\upi="0\UPM19
      \mathchardef\umu="0\UPM16
      \mathchardef\upartial="0\UPM40
      \mathchardef\leqslant="3\AMSa36
      \mathchardef\geqslant="3\AMSa3E

      \let\leq=\leqslant 

    \fi
  \fi
\fi 

\ifnfsstwo
  \DeclareMathAlphabet{\mathbfit}{OT1}{cmr}{bx}{it}
  \SetMathAlphabet\mathbfit{bold}{OT1}{cmr}{bx}{it}
  \DeclareMathAlphabet{\mathbfss}{OT1}{cmss}{bx}{n}
  \SetMathAlphabet\mathbfss{bold}{OT1}{cmss}{bx}{n}
  \ifAMStwofonts
    \ifCUPmtlplainloaded \else
      \DeclareSymbolFont{UPM}{U}{eur}{m}{n}
      \SetSymbolFont{UPM}{bold}{U}{eur}{b}{n}
      \DeclareSymbolFont{AMSa}{U}{msa}{m}{n}
      \DeclareMathSymbol{\upi}{0}{UPM}{"19}
      \DeclareMathSymbol{\umu}{0}{UPM}{"16}
      \DeclareMathSymbol{\upartial}{0}{UPM}{"40}
      \DeclareMathSymbol{\leqslant}{3}{AMSa}{"36}
      \DeclareMathSymbol{\geqslant}{3}{AMSa}{"3E}

      \let\leq=\leqslant 

    \fi
  \fi
\fi 

\ifCUPmtlplainloaded \else
  \ifAMStwofonts \else 
    \def\upi{\pi}
    \def\umu{\mu}
    \def\upartial{\partial}
  \fi
\fi

\title{Simulations and interpretation of the 6~cm MERLIN images of the classical nova V723 Cas}
\author[Heywood \& O'Brien]
{I.~Heywood$^{1,2}$, T.~J.~O'Brien$^2$\\$^1$University of Oxford Astrophysics, Keble Road, Oxford, OX1 3RH, UK\\$^2$University of Manchester, Jodrell Bank Observatory,
Macclesfield, Cheshire SK11 9DL, UK}
\date{Accepted 200x month xx. Received 200x month xx}
\pagerange{\pageref{firstpage}--\pageref{lastpage}}
\pubyear{200x}

\def\LaTeX{L\kern-.36em\raise.3ex\hbox{a}\kern-.15em
    T\kern-.1667em\lower.7ex\hbox{E}\kern-.125emX}

\begin{document}

\label{firstpage}

\maketitle

\begin{abstract}

We compare the predictions of simple models for the radio emission from classical novae with the MERLIN radio observations of nova V723 Cas (Nova Cas 1995). Spherically symmetric and ellipsoidal radiative transfer models are implemented in order to generate synthetic emission maps. These are then convolved with an accurate representation of the $uv$ coverage of MERLIN. The parameters and geometry of the shell model are based on those returned by fitting models to the observed light curve. This allows direct comparison of the model images with the nine 6~cm MERLIN images of V723 Cas.

It is found that the seemingly complex structure (clumping, apparent rotation) evident in the observations can actually be reproduced with a simple spherical emission model. The simulations show that a 24 hour track greatly reduces the instrumental effects and the synthetic radio map is a closer representation of the true (model) sky brightness distribution. It is clear that interferometric arrays with sparse $uv$ coverage (e.g. MERLIN, VLBA) will be more prone to these instrumental effects especially when imaging ring-like objects with time dependent structure variations. A modelling approach such as that adopted here is essential when interpreting observations. 

\end{abstract}


\section{Introduction}

Classical novae are the result of thermonuclear explosions in interacting binary stars. The system consists of a hot white dwarf primary with a main sequence secondary companion. The secondary is thought to fill its Roche lobe and transfers hydrogen-rich material via an accretion disc onto the surface of the white dwarf. When the pressure in the degenerate material at the base of the accreted layer reaches a critical value a thermonuclear runaway takes place (Starrfield 1989), ejecting $\sim$10$^{-5}$--10$^{-4}$~M$_{\odot}$ of material into the interstellar medium at velocities of up to several thousand kilometres per second. This causes an increase in the bolometric luminosity of the system of a few orders of magnitude over a timescale of a few days. 

Radio emission from the nova starts to develop at a later time than the optical. The white dwarf star photoionises the ejecta giving rise to free-free emission at temperatures of approximately 10$^{4}$~K. The evolution of the ejecta is generally rapid, allowing radio interferometric observations to track the expanding shell as it brightens and then fades over favourable timescales (e.g. Taylor et al, 1987; Eyres et al, 1996, 2000, 2005; Heywood et al, 2002, 2005). Both radio and optical images of classical novae often show non-homogeneous, albeit spherical or ellipsoidal structures. A feature common to all MERLIN observations of novae is apparent rotation between epochs. 

In this paper we investigate the structure seen in MERLIN observations of V723 Cas (Nova Cas 1995) using models fitted to the radio light curve (Heywood et al 2005 - hereafter Paper I). The simulations presented in this paper provide an insight into how the radio images that result from interferometer observations relate to the true distribution of the emitting material. Radio images are often cited during discussions of the morphology of nova shells. Although it is common knowledge that interferometer images are an approximation of the true sky brightness distribution, it has not previously been determined just how many of the features which are common to radio observations of novae are due to the interferometer and the aperture synthesis process. The goal for this investigation is to generate synthetic emission maps based on shell models to be subsequently convolved with an accurate $uv$ coverage model of the MERLIN interferometer. 

\begin{figure*}
\begin{center}
\setlength{\unitlength}{1cm}
\begin{picture}(10,21)
\put(-4.0,2.2){\includegraphics{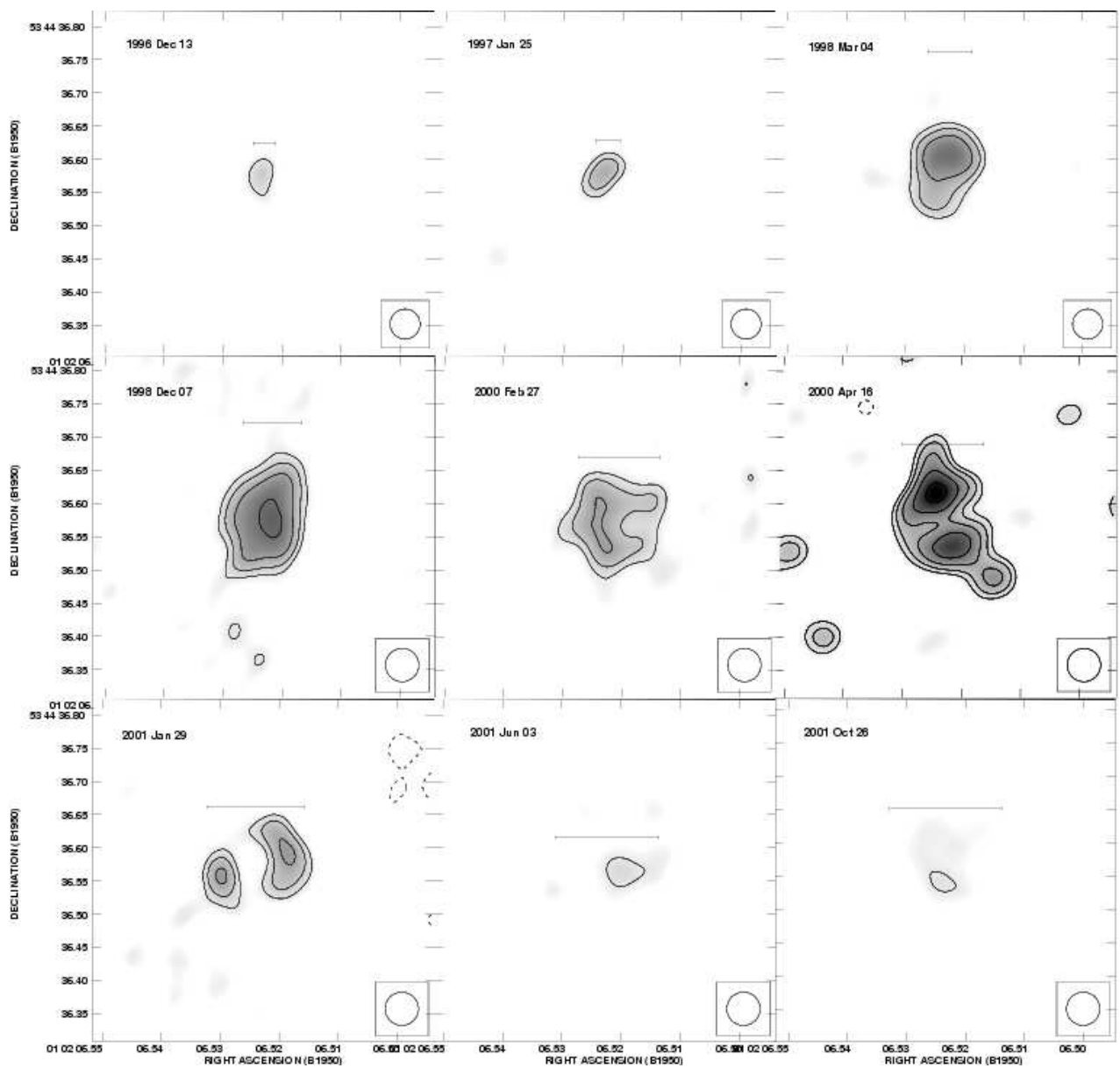}}
\end{picture}
\caption{6~cm MERLIN images of the nova V723 Cas, as published by Heywood et al. (2005), reproduced here for comparison with the simulations. The contour levels are (-3, 3, 3$\sqrt{2}$, 6, 6$\sqrt{2}$, 12, 12$\sqrt{2}$...) $\times$ 0.15~mJy.\label{fig:merlin}}
\end{center}
\end{figure*}

Descriptions of the shell models used are presented in Section \ref{sec:simulations}. A radiative transfer solution is implemented for both a spherically symmetric and an ellipsoidal shell based on the models described in Paper I. This process results in simulated emission maps which are then convolved with a $uv$ model. In all cases this process, described in Section \ref{sec:merlin}, is consistent with the genuine MERLIN observations of V723 Cas presented in Paper I and reproduced in Figure \ref{fig:merlin}. This facilitates direct comparisons between the real and simulated radio images.

\section{Simulations}
\label{sec:simulations}
Model images of the sky brightness distribution are generated by solving the line of sight equations of radiative transfer through a model emitting region. Descriptions of the radiative transfer code and the model shell geometries are presented in this section, together with images showing the predicted brightness distributions. Synthetic interferometer maps are derived by masking the Fourier transform of these images with the $uv$ coverage of the interferometer and then inverting the transform, this process is described in Section \ref{sec:merlin}.

\subsection{Axisymmetric radiative transfer}

\begin{figure}
\centering
\includegraphics[width=\columnwidth]{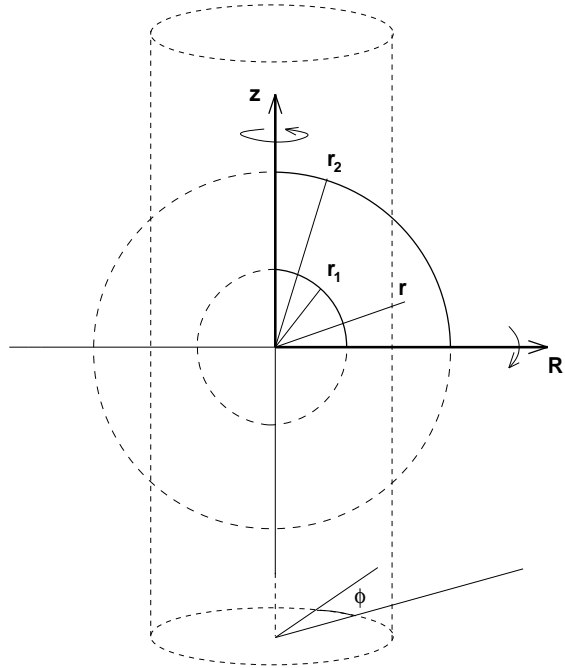}
\caption{Coordinate system for the axisymmetric radiative transfer
code. This figure shows how the shell is defined in the cylindrical
polar coordinates used by the model.}
\label{fig:radcodefig}
\end{figure}

The radiative transfer code uses cylindrical polar coordinates to
define the emitting region as used by, for example, Dougherty et al (2003). The coordinate system is shown in Figure
\ref{fig:radcodefig}. The shell model is constructed by
creating a density function in the plane defined by $R > 0$ and $z >
0$. A 3D shell is created by mirroring the
distribution about the plane defined by $R$ and $\phi$ and rotating
through 360 degrees about the $z$ axis. Thus the quarter circles
indicated by the solid lines in Figure \ref{fig:radcodefig} would define full spherical
regions. The configuration shown in Figure \ref{fig:radcodefig} is representative of the Hubble flow model described in Paper I (and previously by Seaquist \& Palimaka, 1977, and by Hjellming et al, 1979) and is discussed further below.

The density and temperature
are defined in an array of dimensions 256 $\times$ 512 in $R$ and $z$
respectively and when the coordinate transform is applied to define the 3D
emitting region the result is mapped on to a 256 $\times$ 256 pixel image. The
arrays defining the image are chosen to be 2$^{n}$ in size, where $n$
is an integer, so that the Fast Fourier Transform (FFT) technique can be applied
when convolving these images in the $uv$ plane. 

The frequency of the radiation is specified and the radiative transfer
equations are solved along gridded lines of sight through the
shell. The code then calls a Hierarchical Data System (HDS) library which outputs images of
intensity, optical depth and brightness temperature in the STARLINK N-dimensional Data Format (NDF).

An inclination angle for the orientation of the model can also be
specified. This is denoted by $\theta$ and is the angle between the $v$ axis in the plane of the
sky and the $z$ axis of the coordinate system. 

\subsection{Spherically symmetric Hubble flow model}
\label{synhubflow}

The Hubble flow model applied to V723 Cas in Paper I is used as the
basis for the first run of the axisymmetric radiative transfer
code. The values of mass and inner and outer ejection velocity
returned by the best fit routine are used to define the density and the shell geometry for this
model. The inner and outer shell boundaries, and consequently the shell density, are time dependent and emission maps are created with epochs corresponding to the 6~cm MERLIN observations presented in Paper I.

The density function of the Hubble flow model is given by
\begin{equation}
\rho(r) = \frac{M}{4 \pi r^{2}(r_{2} - r_{1})}
\end{equation}
which is a function of $r$,
the radial distance from the centre of the shell, and also of $r_{1}$
and $r_{2}$, the radii of
the inner and outer shell boundaries respectively. Hence when the
shell is defined in the radiative transfer code the density array is filled according to
\begin{equation}
\rho(R,z) = \frac{M}{4 \pi (R^{2} + z^{2}) (r_{2} - r_{1})}
\end{equation}
for $r_{1} \leq r \leq r_{2}$. This is a function of $R$ and $z$ and
the parameters $r_{2}$ and $r_{1}$ which are fixed for a given epoch.
Similarly for $r_{1} \leq r \leq r_{2}$ the temperature is fixed at
17000~K (see Paper I). Outside the shell boundaries the density is assumed to be
zero and the temperature is set at 100~K. The inclination angle is a redundant parameter in the spherical case.

\begin{figure*}
\begin{center}
\setlength{\unitlength}{1cm}
\begin{picture}(10,21)
\put(-3.9,2.8){\includegraphics{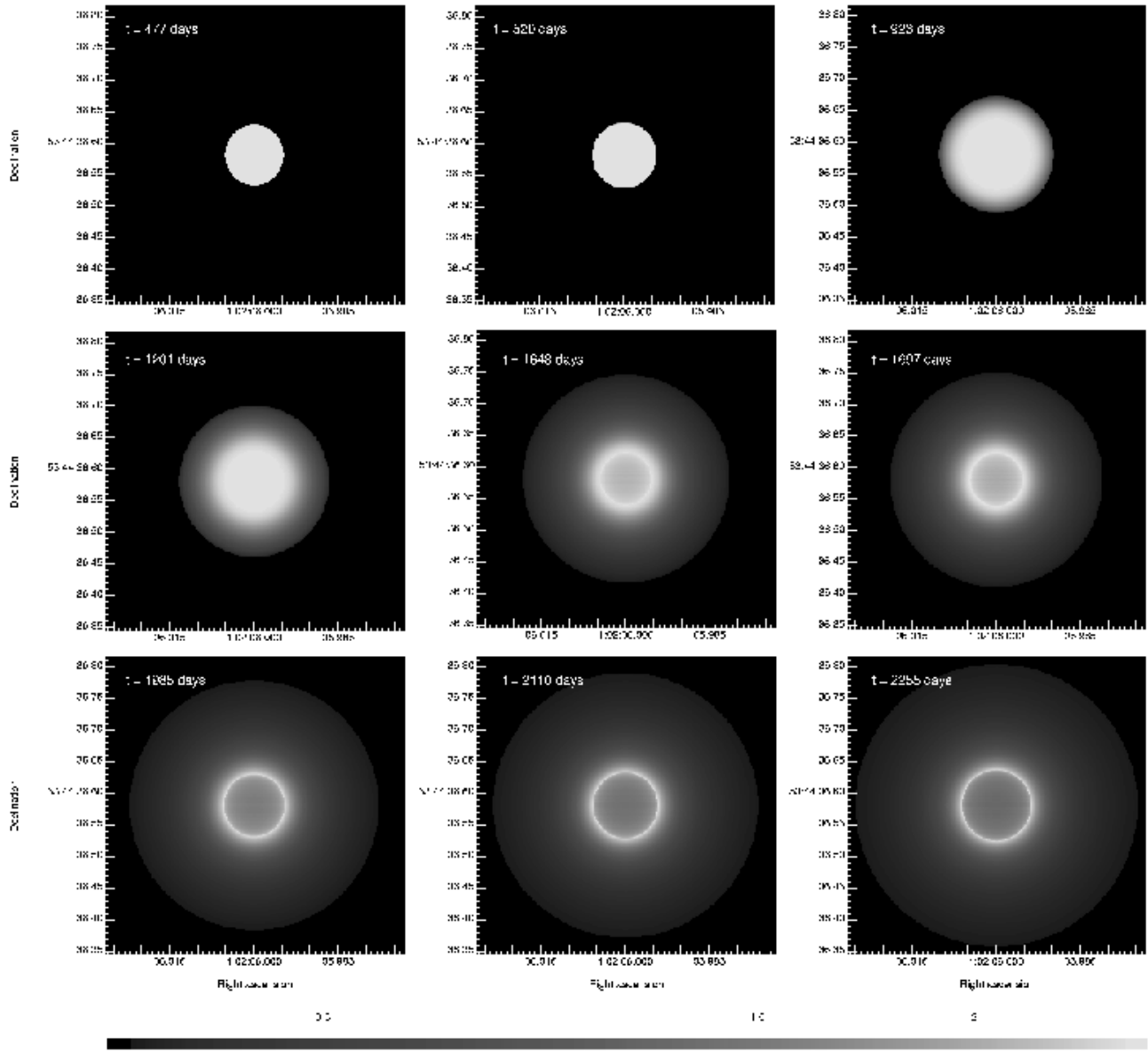}}
\end{picture}
\caption{Model intensity distributions for the Hubble flow model. These maps are calculated at times following outburst shown in each panel, coinciding with the 6~cm MERLIN observations of V723 Cas. The greyscale is in units of mJy. \label{fig:sph_no_uv}}
\end{center}
\end{figure*}

The nine model intensity maps created by this process are presented in
Figure \ref{fig:sph_no_uv} where it can be seen how initially the emission is
entirely optically thick. As the material expands the outer layers
start to become optically thin and eventually the inner boundary of
the shell is revealed as a bright ring.

\subsection{Ellipsoidal shell model}
\label{sec:ellipticals}

Some observations of
classical novae show prolate ellipsoidal shells (e.g. FH Ser, Gill \& O'Brien, 2000) and the
intention here is to determine whether it is possible to distinguish
between a spherical shell or an ellipsoidal shell by studying the radio
images.

\begin{figure}
\centering
\includegraphics*[width=0.9 \columnwidth]{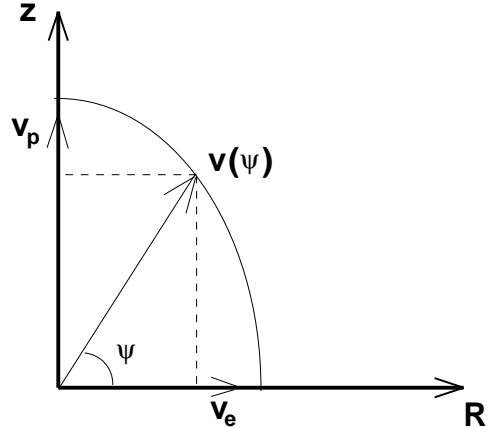}
\caption{Geometry for elliptical shell model.}
\label{ellshell}
\end{figure}

The shell model is based on the spherical Hubble flow model
as described in Section \ref{synhubflow} however some additional
parameters are introduced. 
A polar and equatorial direction are defined which, with reference to
Figure \ref{fig:radcodefig}, are along the $z$ and $R$ axes
respectively. The polar velocity, $v_{p}$, is greater than the equatorial
velocity, $v_{e}$, and the two are related by the factor $\mathcal{P}$
such that $v_{e}$~=~$\mathcal{P}$~$v_{p}$. 
The velocity in these regions is
a function of the angle $\psi$ which is defined as
\begin{equation}
\psi = \tan^{-1}\frac{z}{R}
\end{equation}
such that $v(\psi)$ = $v_{e}$ at $\psi$ = 0  and $v(\psi)$ = $v_{p}$
at $\psi$ = $\pi$/2, see Figure \ref{ellshell}.

The velocity as a function of $\psi$ is governed by the ellipse
equations. In the general case, for an ellipse of semi--major axis $a$,
semi--minor axis $b$ and eccentricity $e$, the radius $r$ as a function of $\theta$,
the angle between the semi--minor and semi--major axes, is given by
\begin{equation}
\label{ellipseradius}
r = \frac{a \sqrt{1 - e^{2}}}{\sqrt{1 - e^{2}~ \cos^{2} \theta}}
\end{equation}
where
\begin{equation}
\label{eccentricity}
e = \frac{\sqrt{a^{2} - b^{2}}}{a}.
\end{equation}
By subtituting Equation \ref{eccentricity} into Equation
\ref{ellipseradius} it can be shown that 
\begin{equation}
r = \frac{ab}{\sqrt{a^{2} - \left(a^{2} - b^{2}\right)\cos^{2}\theta}}
\end{equation}
and the above formula is used to govern how the velocity varies as a function of
$\psi$ in the shell, i.e.
\begin{equation}
\label{ellvel}
v(\psi) = \frac{v_{p}v_{e}}{\sqrt{v_{e}^{2} - (v_{e}^{2} - v_{p}^{2})\cos^{2}\psi}}.
\end{equation}

The elliptical shell models are again constructed based on the best
fit parameters of V723 Cas as returned by the light curve fitting in Paper I. The best fitting inner and outer shell velocities are taken
to be the velocities of the inner and outer shell boundaries at the
pole. The shell mass is taken to be the best fitting shell mass from the spherical model. The value of  $\mathcal{P}$ is
arbitrarily chosen to be 0.6 and the inclination of the shell to the plane of the sky is taken
to be zero degrees i.e. the shell is viewed `side-on'.

As with the spherically symmetric model, nine simulated maps are calculated to correspond to the nine epochs of
6~cm MERLIN observations. Equation \ref{ellvel} is used to determine
$v_{1}(\psi)$ and $v_{2}(\psi)$, 
the velocities of the inner and outer shell boundaries as a function
of $\psi$, and these velocities are then
used to calculate $r_{1}(\psi)$ and $r_{2}(\psi)$, the radii of the
inner and outer shell boundaries for the specific epoch. The density
function 
\begin{equation}
\rho(R,z) = \frac{M}{4 \pi (R^{2} + z^{2}) (r_{2} - r_{1})}
\end{equation}
is then applied to populate the density array in $R$ and $z$. Once
again the temperature is set at 17000~K within the shell boundaries
and is assumed to be 100~K elsewhere.

\begin{figure*}
\begin{center}
\setlength{\unitlength}{1cm}
\begin{picture}(10,21)
\put(-3.9,2.79){\includegraphics{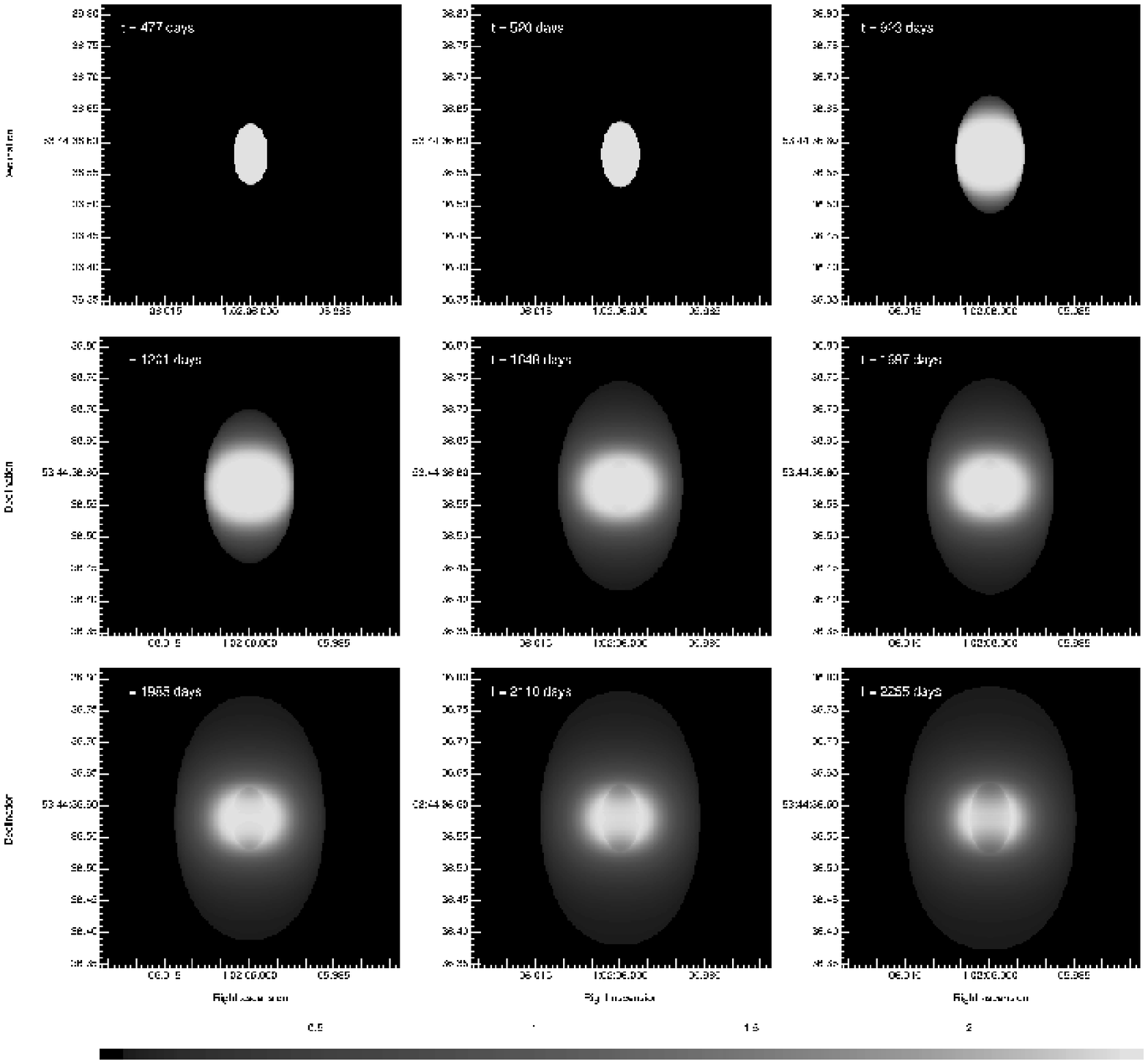}}
\end{picture}
\caption{Model intensity distributions for the elliptical shell model, see Fig.\ \ref{fig:sph_no_uv}. \label{fig:ell_no_uv}}
\end{center}
\end{figure*}

The nine model intensity distribution maps are presented in Figure
\ref{fig:ell_no_uv}. As with the spherical case the emission is initially
optically thick before expansion causes the optical depth to decrease
and the inner shell boundary is revealed. An interesting feature to
note is the `waist' of brighter emission in the equatorial region. This is due to the inverse
dependence on the quantity $r_{2} - r_{1}$ in the density
equation. This is smaller for small values of
$\psi$ than it is for values of $\psi$ close to $\pi / 2$ radians giving
rise to a density enhancement near the equator of the shell. 

\section{Simulated MERLIN images}
\label{sec:merlin}

Once the model images have been calculated the image files have to be correctly formatted before they can be imported into the National Radio Astronomy Observatory's Astronomical 
Image Processing System (AIPS; {\tt http://www.aoc.nrao.edu/aips/}) which is used to generate the synthetic MERLIN images. A World Coordinate System, frequency and Stokes parameter axes are added.
The AIPS task UVCON\footnote{Recently a bug was discovered in the AIPS task UVCON whereby in certain cases visibilities in the simulated data set would be recorded
with telescopes at negative elevations, and then conversely visibilities would be missing from high elevations where they should be present (L. Kogan, private communication).

When 
this problem came to light the simulations in this paper were naturally re-examined to check for this problem. This footnote is merely to confirm that the results presented are not affected.
The declination of V723 Cas coupled with the latitudes of the stations in the MERLIN array result in positive elevations throughout, even for a 24 hour track.} reads the properties of the array from an antenna file and determines the $uv$ coverage of the interferometer. The visibilities are then calculated from a gridded FFT of a model image. Since the models were defined by fits to the light curve, the flux density is very close to that measured in the actual observations. Gaussian noise is added to the data, the level of which is calculated by UVCON based on integration times and the total length of the track.  The $uv$ data produced by UVCON are deconvolved, cleaned and imaged using IMAGR. 
In all cases where possible, the imaging process has been consistent with the methods used to produce the 6~cm MERLIN maps shown in Figure \ref{fig:merlin}. The cleaning process was terminated at the first negative clean component and a circular restoring beam of 50~mas was used. 

\begin{figure*}
\begin{center}
\setlength{\unitlength}{1cm}
\begin{picture}(10,21)
\put(-3.2,3.2){\includegraphics{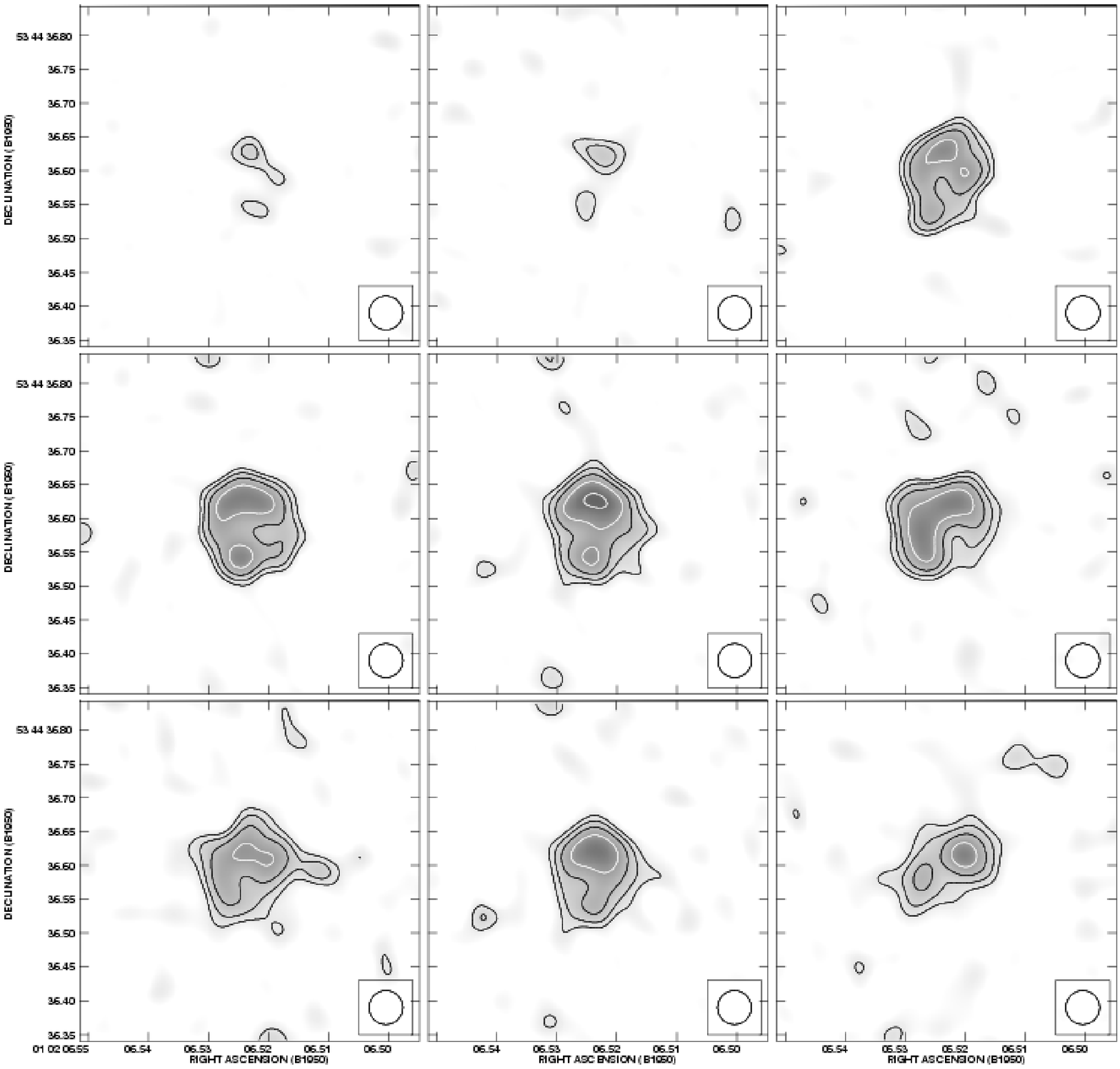}}
\end{picture}
\caption{Simulated radio maps for the elliptical shell model based on a 12 hour track. The contour levels are (-3, 3, 3$\sqrt{2}$, 6, 6$\sqrt{2}$, 12, 12$\sqrt{2}$...) $\times$ 0.15~mJy.\label{fig:ellnine}}
\end{center}
\end{figure*}

Figure \ref{fig:sphnine} shows the simulated MERLIN images based on the spherically symmetric Hubble flow model. These images are based on a 12 hour track which is of similar length to the original MERLIN observations. The images in Figure \ref{fig:24nine} take the same model intensity
distributions as those used to generate Figure \ref{fig:sphnine} except the
simulated $uv$ coverage is based on a 24 hour track instead of a
12 hour track. Figure \ref{fig:ellnine} shows the model MERLIN observations of the elliptical shell model based on a 12 hour track, and finally Figure \ref{fig:24ell} shows the same model with a 24 hour track.
In each 3 $\times$ 3 grid, left to right, top to bottom, the images correspond to the December 1996, January 1997, March 1998, December 1998, February 2000, April 2000, January 2001, June 2001 and October 2001 MERLIN observations of V723 Cas. Contour levels are (-3, 3, 3$\sqrt{2}$, 6, 6$\sqrt{2}$, 12, 12$\sqrt{2}$...) $\times$ 0.15~mJy for each image, as quoted in the captions.

\section{Discussion}

\begin{figure*}
\begin{center}
\setlength{\unitlength}{1cm}
\begin{picture}(10,21)
\put(-4.4,-1.5){\includegraphics{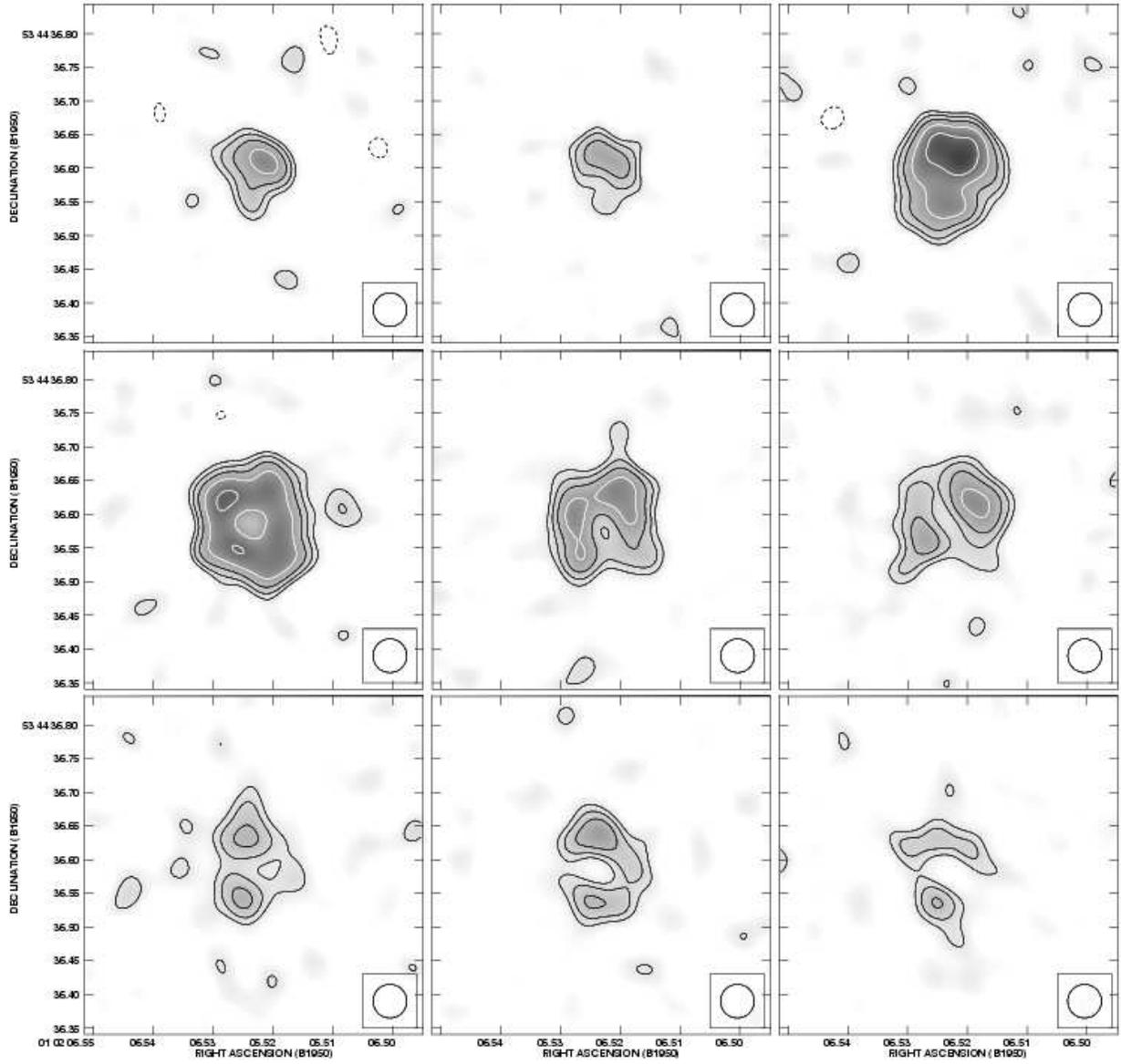}}
\end{picture}
\caption{Simulated radio images for the spherically symmetric Hubble flow model based on a 12 hour track. The contour levels are (-3, 3, 3$\sqrt{2}$, 6, 6$\sqrt{2}$, 12, 12$\sqrt{2}$...) $\times$ 0.15~mJy.\label{fig:sphnine}}
\end{center}
\end{figure*}

Comparison of Figure \ref{fig:sphnine} to the MERLIN images presented in Figure \ref{fig:merlin} 
reveals that even for this simple spherical shell model the similarities between
the simulated radio observations and the real MERLIN observations are striking. Initially the source is
point-like, not much bigger than the restoring beam. The source
expands and as the turn over into optically thin emission begins apparent
structure in the shell becomes evident. The source becomes ring-like
then divides into the double-peaked structure which is common to all
multi-epoch MERLIN observations of classical novae. 

\begin{figure*}
\begin{center}
\setlength{\unitlength}{1cm}
\begin{picture}(10,21)
\put(-4.2,-1.93){\includegraphics{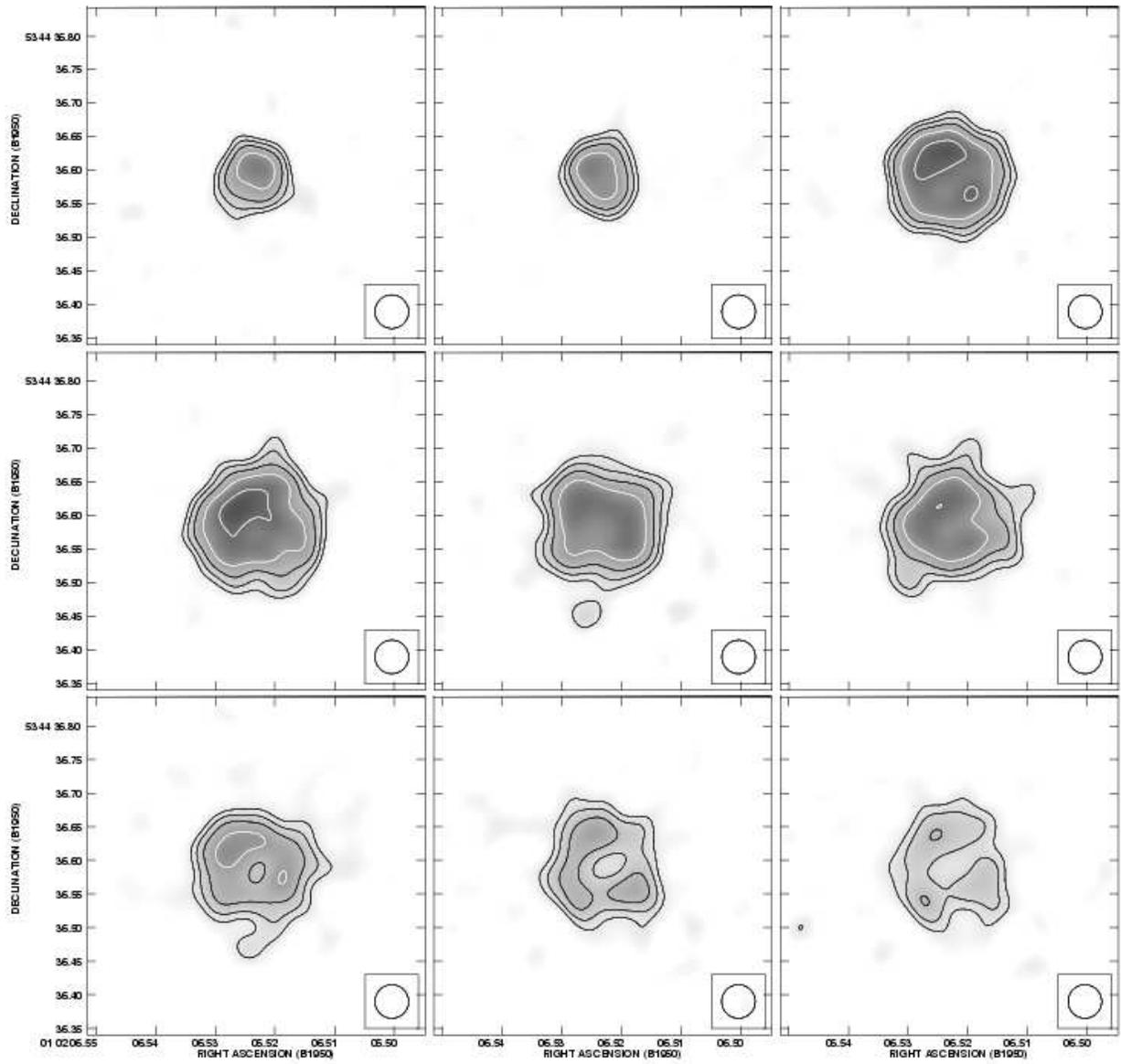}}
\end{picture}
\caption{Simulated radio maps for a 24 hour track of the spherical Hubble flow model. The contour levels are (-3, 3, 3$\sqrt{2}$, 6, 6$\sqrt{2}$, 12, 12$\sqrt{2}$...) $\times$ 0.15~mJy.\label{fig:24nine}}
\end{center}
\end{figure*}

\begin{figure*}
\begin{center}
\setlength{\unitlength}{1cm}
\begin{picture}(10,21)
\put(-3.2,2.7){\includegraphics{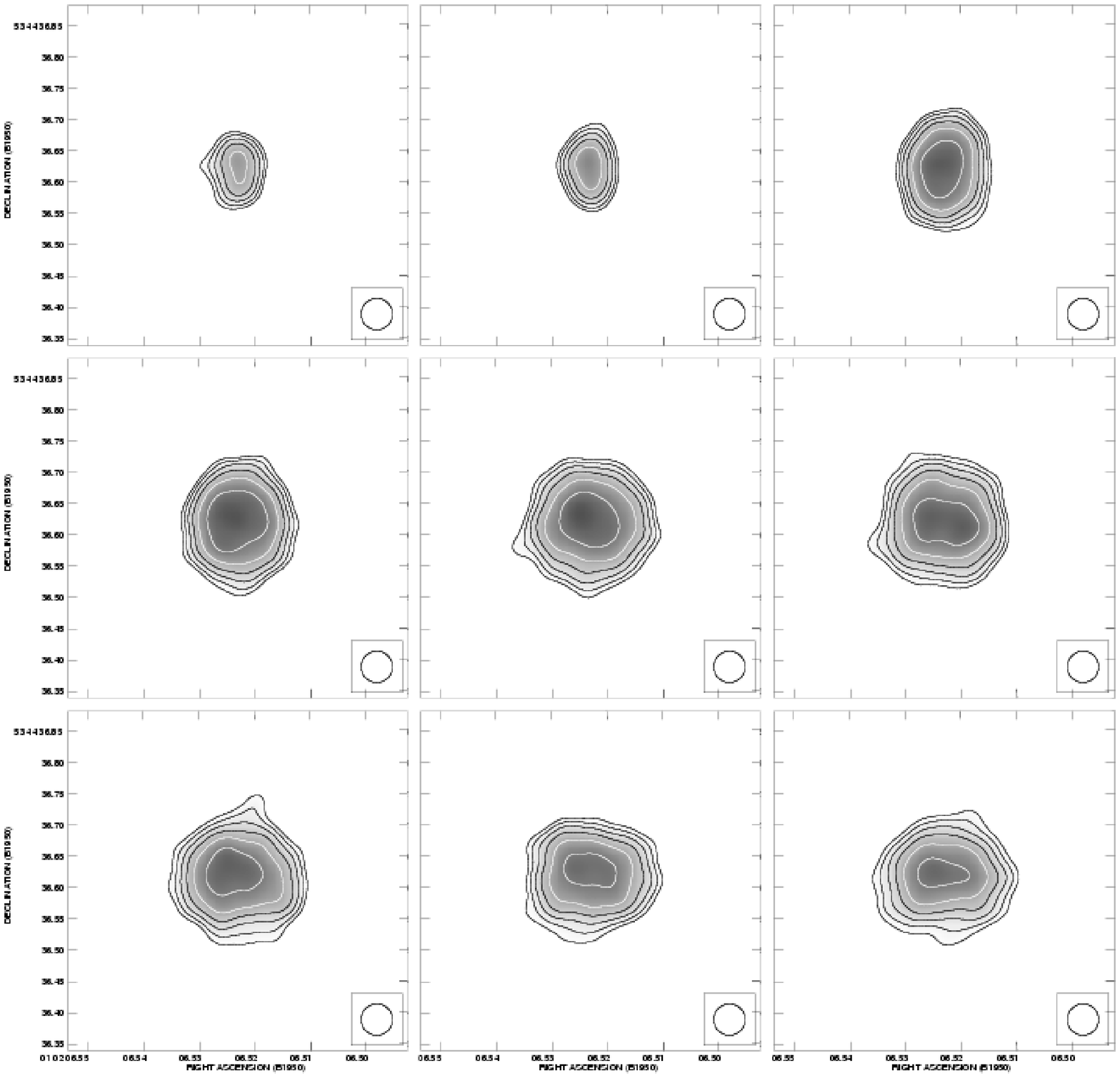}}
\end{picture}
\caption{Simulated radio maps for the elliptical shell model based on a 24 hour track. The contour levels are (-3, 3, 3$\sqrt{2}$, 6, 6$\sqrt{2}$, 12, 12$\sqrt{2}$...) $\times$ 0.15~mJy.\label{fig:24ell}}
\end{center}
\end{figure*}

The most intriguing feature of these simulated maps is the switching of the
alignment of the dominant peaks in the map, giving the impression of rotation.
In Paper I we speculated whether this was due to optical depth effects, with higher
density regions of the shell remaining optically thick for longer and appearing as peaks of radio emission as the
photosphere recedes. Here this explanation can be ruled out as the mass
distribution supplied by the model is spherical with a smooth radial density profile.

Comparing Figure \ref{fig:24nine} to Figure \ref{fig:sphnine},
shows that for a 24 hour observation
the radio emission in the synthetic maps is more circularly symmetric  and
consquently more faithful to the model images. The effect whereby
the alignment of the dominant peaks in emission appears to rotate
between epochs is also almost completely eradicated. The increase in $uv$
coverage provided by the 24 hour observation must be the reason for
this, providing both an improved signal to noise ratio and a drastically altered sidelobe structure in the beam. The beam shape and $uv$ coverage are discussed in Section \ref{sec:beamsanduv}. Table \ref{tab:rms} lists the root mean square values of the background noise for both 12 and 24 hour tracks of the simulated spherical Hubble flow model.

\begin{table}
\centering
\caption{Comparison of the background rms values in mJy/beam for both the 12 and 24 hour simulations of the spherical Hubble flow model.\label{tab:rms}}
\begin{tabular}{lll}
Epoch & 12 hour track & 24 hour track\\
\hline
Dec 96 & 0.169 & 0.124 \\
Jan 97 & 0.129 & 0.119 \\
Mar 98 & 0.177 & 0.137 \\
Dec 98 & 0.185 & 0.141 \\
Feb 00 & 0.141 & 0.134 \\
Apr 00 & 0.137 & 0.133 \\
Jan 01 & 0.136 & 0.124 \\
Jun 01 & 0.131 & 0.117 \\
Oct 01 & 0.128 & 0.112 \\
\hline
\end{tabular}
\end{table}

The simulated radio maps for the ellipsoidal shell model reveal a
north--south bias with a double peaked source being present for
the first five epochs. This gives way to a `partial shell' structure
which in turn is replaced by the east--west aligned peaks in the final
map. Although a stronger north--south alignment of the emission would
be expected from this ellipsoidal shell model the evolution of the
radio emission over time displays many similiarities to the spherical
case. Both initially develop a north--south bias which gives way to an
east--west alignment and although this process occurs later for the
ellipsoidal shell it is unlikely that a 12 hour track can be used
to distinguish between an ellipsoidal or a spherical shell for objects
of this nature.

In order to determine whether a 24 hour track would produce a distinction between ellipsoidal and spherical shells the simulations presented in Figure \ref{fig:24ell} can be examined. For the optically thick stage the MERLIN images accurately represent the simulated brightness distributions but for subsequent epochs the images detect only the brighter east-west aligned emission around the inner shell boundary. The faint, extended structure along the major axis of the shell is below detectability. With a greater signal to noise ratio, or alternatively an object brighter than V723 Cas, it is likely that observations based on a 24 hour track could be used to examine the eccentricity of an ellipsoidal shell.

\subsection{Comparing the beams and the $uv$ coverage}
\label{sec:beamsanduv}

\begin{figure}
\centering
\includegraphics[width= 0.9 \columnwidth]{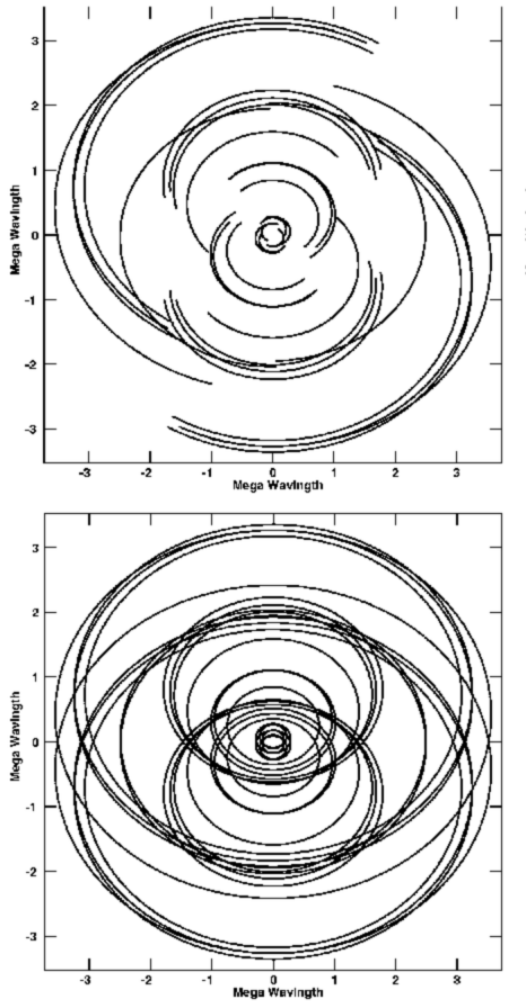}
\caption{Plots of the simulated $uv$ coverage for a 12 hour (upper panel) and 24 hour (lower panel) track.\label{fig:12_24_UV}}
\end{figure}

\begin{figure}
\centering
\includegraphics[width= 0.75 \columnwidth]{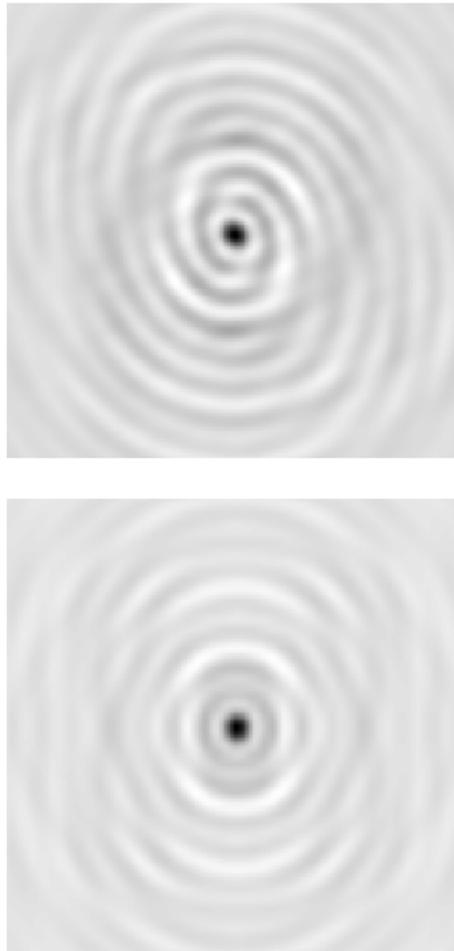}
\caption{Plots of the beam structure for the simulated 12 hour
(upper panel) and 24 hour (lower panel) observations. The beam structure is shown on the same scale as the simulated radio images.}
\label{fig:12_24_beam}
\end{figure}

Figure \ref{fig:12_24_UV} shows the improved $uv$ coverage of the 24 hour track compared to the 12 hour track.
Plots of the beam shape for both 12 and 24 hour observations are
presented in Figure \ref{fig:12_24_beam}. Immediately apparent is the effect that the increased $uv$
coverage has on the sidelobes of the beam.  The sidelobe structure in the 24 hour observation
completely surrounds the central maximum and only deviates from
circularity due to the declination of the source. The beam of the 12 hour
simulation has $S$-shaped sidelobe structure. As the source
expands through the beam different parts of the shell will be buried in the sidelobes. 
The non-circular beam shape associated with the 12 hour track could be a possible cause for the
apparent rotation of the nova shell between different epochs. Since this effect is generally not an issue with 
observations involving strong sources on larger scales (e.g. radio galaxies) it is likely that these effects will 
only be present when observing objects with low surface brightness objects and small angular sizes, and would 
probably pass unnoticed when observing systems with less favourable dynamic timescales. Additionally a 24 hour track 
provides higher fidelity images due to more thorough sampling of the $uv$ plane and an improved signal to noise ratio.

\subsection{Rings and discs}

\begin{figure}
\centering
\includegraphics[width= \columnwidth]{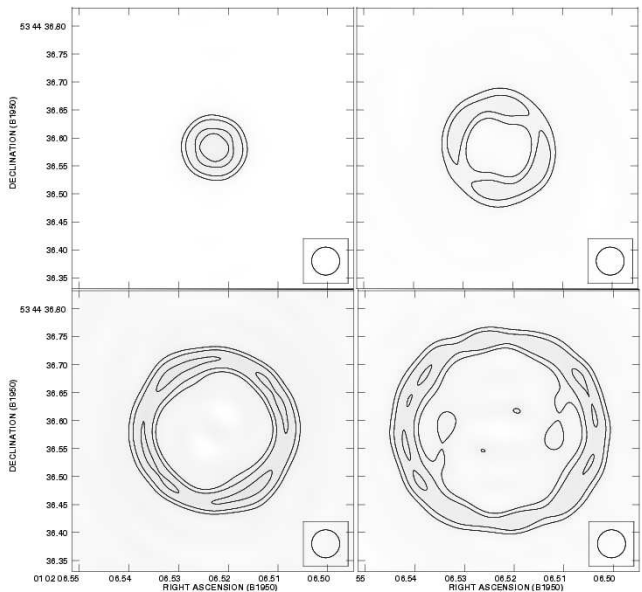}
\caption{Synthetic 12 hour MERLIN observations of an expanding ring.\label{fig:rings}}
\end{figure}

\begin{figure}
\centering
\includegraphics[width= \columnwidth]{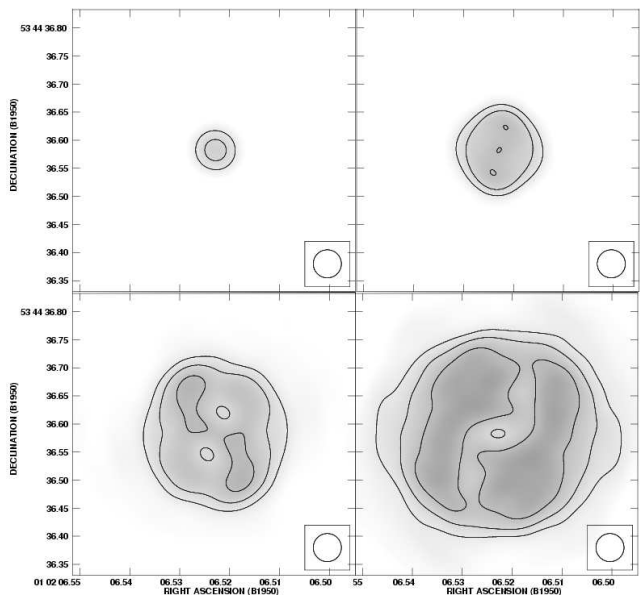}
\caption{Synthetic 12 hour MERLIN observations of an expanding disc.\label{fig:discs}}
\end{figure}

The Fourier transform process can give rise to unexpected artefacts in the final image. In this section we briefly investigate the possibility that the transition from a disc shaped structure to something that is ring-like could be responsible for unusual effects in the final radio images. Examination of Figure \ref{fig:sph_no_uv} shows how the model emitting region develops with time. Initially the source is optically thick and resembles a disc on the sky, with the flux density at this stage governed only by the angular size and the brightness temperature. The highest contour level also drifts around the field as the shell expands. Later a transition to optically thin emission occurs and the ring-like inner shell boundary is revealed. In order to assess the outcome of this transition on the final radio images, simulated MERLIN observations of an expanding ring are presented for comparison with those of an expanding disc. These are presented in Figures \ref{fig:rings} and \ref{fig:discs}. 

The model emitting region in each case has angular size scales and flux densities comparable with that of the shell of V723 Cas. As the intention here is to solely investigate the effect of the deconvolution process the final simulated data do not have Gaussian noise added. Note the differing contour levels for the ring and disc maps. The disc simulation is plotted with levels of (-3, 3,
3$\sqrt{2}$, 6, 6$\sqrt{2}$, 12, 12$\sqrt{2}$...) $\times$ 0.15~mJy as
with the real MERLIN maps and the simulations presented in this paper. Since the simulated ring structure is much
fainter the contour levels are reduced to (-3, 3, 3$\sqrt{2}$, 6, 6$\sqrt{2}$, 12, 12$\sqrt{2}$...) $\times$ 0.03~mJy. 
The disc simulation is brighter simply because there is more emitting material.

Examination of Figures \ref{fig:rings} and \ref{fig:discs} reveals that even
these simple zero noise models exhibit features such as apparent rotation 
and peaks dividing as the emitting region expands. These
effects are seen in both the ring and the disc model meaning that it
is unlikely that they appear in the real observations due to the nova
undergoing the transition from disc-like to ring-like. This is further
evidence that these effects are due to the limited $uv$ coverage 
and are not related to the changing optical depth or other underlying properties of the 
source.

\section{Conclusions}

These results have shown that extreme care must be taken when using radio images alone to interpret the morphology of classical nova shells such as that of V723 Cas. The (comparatively) low radio surface brightness of a nova shell coupled with its low angular size gives makes it prone to instrumental effects which may be mistaken for unusual shell morphology and motion (e.g. clumping, apparent rotation). However these effects can be mitigated by using observations with longer tracks which both improves the $uv$ coverage and increases the signal to noise ratio.
When using radio imaging to track structure in novae and similar objects, coupling of simulations such as those presented here with observations would provide a more thorough insight into the underlying source structure. 

\begin{figure}
\centering
\includegraphics[width= \columnwidth]{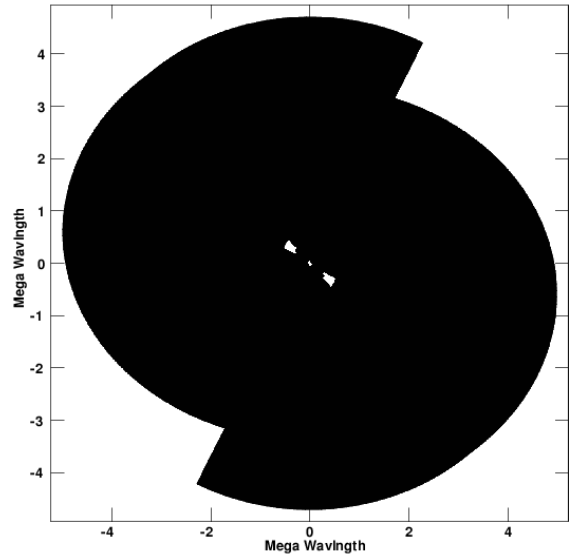}
\caption{Simulated $uv$ coverage of a 12 hour track using the $e$-MERLIN 5~GHz contiguous band.\label{fig:emerlin_uv}}
\end{figure}

\begin{figure}
\centering
\includegraphics[width= 0.8 \columnwidth]{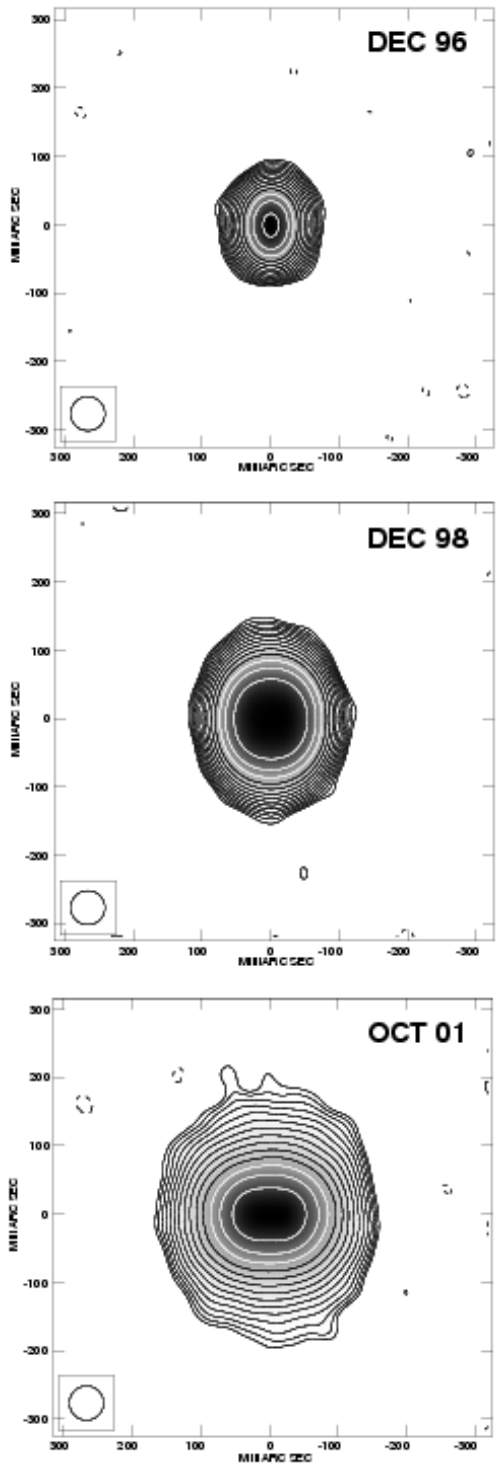}
\caption{Preliminary $e$-MERLIN simulations of the December 1996, December 1998 and October 2001 epochs of the elliptical shell model.\label{fig:emerlin_sims}}
\end{figure}

However, optical imaging of novae reveal that the emission is far from homogeneous and exhibits equatorial and polar enhancements (e.g. FH Ser, Gill \& O'Brien, 2000), clumping of material and even filamentary structures (e.g. GK Per, Slavin et al, 1995). Thus there is need for spherical models to be refined and it will be interesting to see how novae are observed by the next generation of interferometers such as $e$-MERLIN. With the vastly improved $uv$ plane sampling of $e$-MERLIN exploiting multifrequency synthesis, Figure \ref{fig:emerlin_uv}, one would not expect the instrumental effects documented in this paper to be present. Preliminary  simulations shown in Figure \ref{fig:emerlin_sims} show three epochs of the elliptical shell model described in Section \ref{sec:ellipticals}. Immediately apparent is the north-south aligned, extended, low brightness emission which was below detectability levels in the simulations presented in Section \ref{sec:merlin}. The root mean square background noise levels in the $e$-MERLIN simulations are $\sim$5~$\mu$Jy.  The $uv$ plane sampling is consistent with that of $e$-MERLIN observing using a 5~GHz contiguous band, i.e. 1000 frequency channels between 2~GHz and 5~GHz. The drawback here is that these simulations treat the source as being spectrally flat, which is obviously a poor assumption over such a wide frequency range. Nonetheless they show that the $e$-MERLIN view of sources of this nature will be much more faithful to the true source distribution due to its vastly improved $uv$ coverage and sensitivity.

\section*{Acknowledgements}

IH wishes to thank the Particle Physics and Astronomy Research Council (PPARC). MERLIN is operated as a UK National Facility by the University of Manchester, 
Jodrell Bank Observatory, on behalf of PPARC. IH also thanks Simon Garrington for his valuable input. This research has made use of NASA's Astrophysics Data System. The authors wish
to thank the reviewer, Amy Mioduszewski, for her constructive criticisms.

\bsp 

\label{lastpage}

\end{document}